\documentclass[12pt]{article}
\let\section=\subsection     \let\subsection=\subsubsection                
\usepackage{graphicx}

\begin{document}
\begin{center}
   {\large \bf \boldmath 
      High-$p_T$ Neutral Pion Production\\[1mm]
      in Heavy Ion Collisions at SPS and RHIC \unboldmath} \\[5mm]
      K.~Reygers for the WA98 and the PHENIX collaboration\\[5mm]
      {\small \it  Institut f\"ur Kernphysik der Universit\"at M\"unster\\
      Wilhelm-Klemm-Str.~9, D-48149 M\"unster, Germany \\[8mm] }
\end{center}

\begin{abstract}\noindent
Transverse momentum spectra for neutral pions  
have been measured by the WA98 experiment in $\sqrt{s_{nn}}$=17.3~GeV Pb+Pb 
collisions at the CERN SPS and by the PHENIX experiment in
$\sqrt{s_{nn}}$=130~GeV Au+Au collisions at RHIC. The neutral pion yields in
central collisions for both reaction systems are compared to scaled 
transverse momentum spectra for nucleon-nucleon reactions at the 
respective energy.
At SPS energies neutral pion production at high $p_T$ is enhanced compared 
to the n+n reference while at RHIC a significant suppression of high-$p_T$ 
neutral pions is observed.
\end{abstract}

\section{Introduction}
\label{sec:intro}
In heavy ion reactions particle production at high transverse momentum
is expected to result from parton scatterings with large momentum
transfer. In n+n reactions these hard scatterings dominate particle 
production above $p_T\approx 2$~GeV/$c$ \cite{owens}. Hard scattering 
in heavy ion collisions will occur in the early stage of the 
reaction, well before a quark-gluon plasma is expected to form.
Thus, fast partons produced in hard scatterings interact with the
extended medium that is subsequently produced in a heavy ion reaction.
The interaction of partons with hot and dense nuclear matter 
leads to an energy loss of the partons. This parton energy loss is 
considered as a potential signature of a quark-gluon plasma formation 
since the energy loss in a medium of deconfined quarks and gluons is 
expected to be higher than in hadronic matter \cite{baier}. 

An observable consequence of parton energy loss is the suppression
of hadron production at large transverse momenta relative to 
a baseline expectation in the absence of energy loss. The construction
of this baseline necessarily relies on model assumption. For the description
of hard scattering processes the nucleons of the two colliding nuclei
can be approximated as incoherent superpositions of partons. In this
picture the particle yields in a heavy ion reaction are expected to scale 
with the number $N_{coll}$ of inelastic nucleon-nucleon collisions, since
the cross-section for hard processes is small. Thus, the results on 
neutral pion production are presented in terms of the following ratio 
of invariant yields $E\,d^3N/dp^3$:
\begin{equation}
\label{eq:raa}
R_{AA} = \frac{E\frac{d^3N}{dp^3}(p_T)|_{AA} \, / \, N_{coll}}
              {E\frac{d^3N}{dp^3}(p_T)|_{nn}}
\end{equation} 
This ratio is usually denoted as the nuclear modification factor.
In the absence of nuclear modifications to hard scattering 
$R_{AA}$ will be unity.

Calculations based on perturbative QCD are a second possibility
for the construction of a baseline expectation. These calculations
usually also phenomenologically model the multiple initial scatterings
of the partons in a A+A reaction. By just comparing the hadron yield
in A+A reaction with the scaled n+n reference, as in equation~\ref{eq:raa}, 
the effect of multiple parton scattering is not taken into account.
Multiple parton scattering can explain the enhancement of high-$p_T$
hadron production in p+A reactions relative to a $N_{coll}$-scaled 
n+n reference. This anomalous enhancement is usually referred to as 
the Cronin-effect. In order to fix the strength of initial parton 
scattering in the modeling of A+A reactions data for p+A reactions
at the same energy are very important.

\section{Neutral Pion Spectra at SPS and RHIC}
\label{sec:pions}
\begin{figure}
   \centerline{\includegraphics[height=7cm]{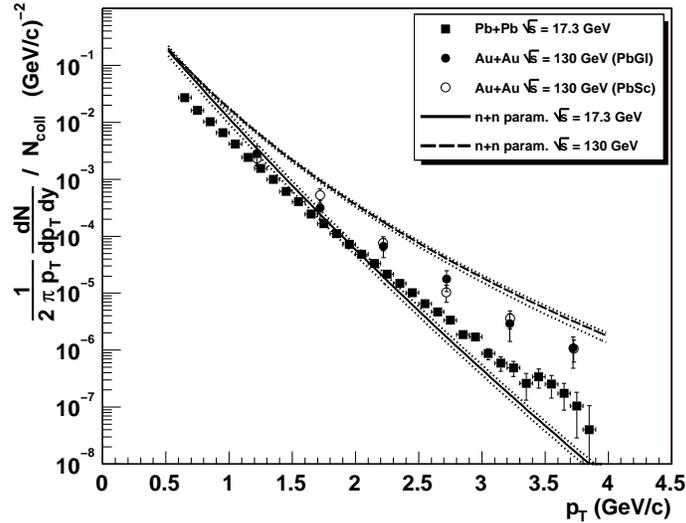}}
   \caption{\footnotesize
            Neutral pion transverse momentum spectra in central Pb+Pb reactions
            at the CERN SPS and in central Au+Au reactions at RHIC. The yields 
            are normalized to the respective number of binary nucleon-nucleon
            collisions. The data are compared to parameterizations of 
            neutral pion spectra in nucleon-nucleon collisions.
            The estimated uncertainties of these parameterizations are 
            indicated by the dotted lines.}
   \label{fig:pi0cen}
\end{figure}

The WA98 experiment at the CERN SPS and the PHENIX experiment at RHIC
reconstruct neutral pions on a statistical basis via their decay into
two photons.  The photon detector LEDA of the WA98 experiment consists
of 10.080 lead glass modules and covers the pseudorapidity interval
$2.3 < \eta < 3.0$. After the last WA98 data taking in 1996 the WA98
lead glass detector was brought from CERN to the Brookhaven National
Laboratory and is now part of the PHENIX experiment. PHENIX consists
of two separate electromagnetic calorimeters: the lead glass detector
(PbGl) and a lead-scintillator (PbSc) sampling calorimeter. Both
detectors cover the pseudorapidity interval $-0.35 < \eta <0.35$. The
PbGl and PbSc data were analyzed separately. Figure~\ref{fig:pi0cen}
shows the agreement of the final $\pi^0$ spectra
\cite{phenix_highpt}.

The number $N_{coll}$ of inelastic binary nucleon-nucleon collisions 
in WA98 and in PHENIX is determined within a Glauber model framework
\cite{wa98_scal,phenix_nch}. 
The Glauber-model is based on a purely geometric picture of a nucleus-nucleus 
collision: Nucleons travel on straight-line trajectories and a collision 
between two nucleons takes place if their distance in the plane transverse 
to the beam axis is smaller than a certain value given by the inelastic 
nucleon-nucleon cross section. The inelastic nucleon-nucleon cross section
increases from $\sigma_{nn} \approx 30$~mb at CERN SPS energies 
($\sqrt{s}=17.3$~GeV) to $\sigma_{nn} \approx 40$~mb at the RHIC energy
of $\sqrt{s}=130$~GeV.

The neutral pion yields in central Pb+Pb collisions at 
$\sqrt{s_{nn}}=17.3$~GeV and in central Au+Au collisions at 
$\sqrt{s_{nn}}=130$~GeV are shown in Figure~\ref{fig:pi0cen}
\cite{wa98_pi0, phenix_highpt}. 
The yields are normalized to $N_{coll}^{WA98} = 651 \pm 65$ and 
$N_{coll}^{PHENIX} = 905 \pm 96$, respectively. The measured 
$\pi^0$ spectra are compared to results for nucleon-nucleon reactions. 
The n+n references is a parameterization that is based on an interpolation 
of existing data to the respective center-of-mass energy 
\cite{wa98_pi0,phenix_highpt}. One observes that the $\pi^0$ spectrum 
in n+n reactions at the RHIC energy is significantly flatter than the 
corresponding spectrum at the CERN SPS energy. The curvature of the 
$\pi^0$ spectrum in A+A reactions, however, changes only moderately when
going from SPS to RHIC.

\begin{figure}
   \centerline{\includegraphics[height=7cm]{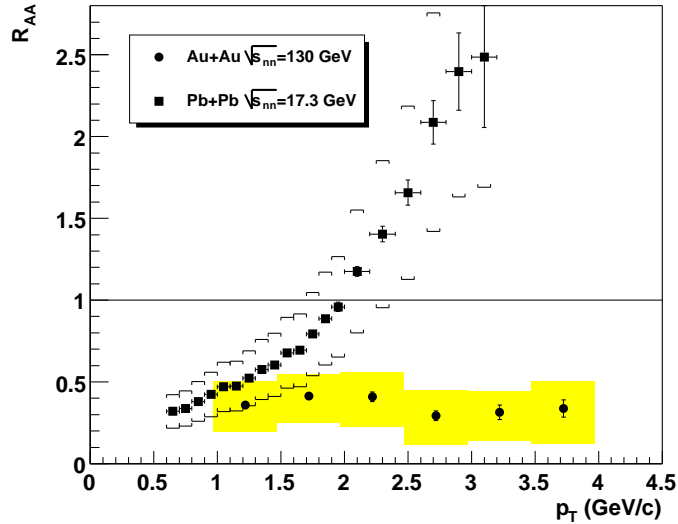}}
   \caption{\footnotesize
            The ratio $R_{AA}$ as defined in equation~\ref{eq:raa} 
            for neutral pion spectra at CERN SPS and RHIC energy. 
            For the WA98 data points the error due to the uncertainty 
            of the n+n reference and due to the uncertainty of $N_{coll}$
            is indicated by brackets. 
            The errors bars of the PHENIX data points indicate the 
            statistical errors. The systematical errors, indicated by
            shaded areas, include the uncertainties of the n+n reference
            and of $N_{coll}$. 
            }
   \label{fig:cen_pp}
\end{figure}

Figure~\ref{fig:cen_pp} shows the ratio $R_{AA}$ as defined in 
equation~\ref{eq:raa}. The PHENIX PbGl and PbSc results were 
averaged for this plot. A striking difference between the results
for the SPS and RHIC energy can be seen. The ratio $R_{AA}$ for 
$\sqrt{s_{nn}}=17.3$~GeV is qualitatively in line with expectations
from the Cronin effect: $R_{AA}$ increases with $p_T$ and the neutral pion 
yield increases stronger than $N_{coll}$ above $p_T \approx 2$~GeV/$c$.
At $\sqrt{s_{nn}}=130$~GeV, however, $R_{AA}$ is basically flat.
Moreover, $R_{AA}$ is significantly below unity over the entire $p_T$ range.
Thus, high-$p_T$ neutral pion production at RHIC energies is suppressed
compared to the $N_{coll}$-scaled n+n reference.    

\begin{figure}
   \centerline{\includegraphics[height=7cm]{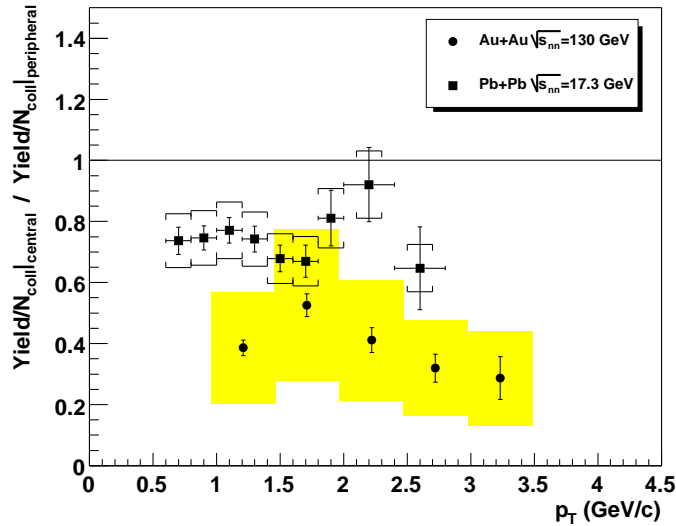}}
   \caption{\footnotesize
            Ratio of central and peripheral $\pi^0$ spectra 
            at SPS and RHIC energy. Each spectrum is normalized 
            to the number of n+n collisions.}
   \label{fig:cen_per} 
\end{figure}

One can now replace the n+n reference by the $N_{coll}$-normalized
$\pi^0$ spectrum measured in peripheral A+A collisions. This is depicted
in Figure~\ref{fig:cen_per}. The number of binary nucleon-nucleon collisions
for the peripheral samples are $N_{coll}^{WA98}=30 \pm 5$ and 
$N_{coll}^{PHENIX}=20 \pm 6$. Figure~\ref{fig:cen_per} indicates that 
the $N_{coll}$ normalized peripheral $\pi^0$ spectrum at RHIC is very
similar to the respective n+n spectrum. In contrast, the comparison
of the SPS results in Figure~\ref{fig:cen_pp} and Figure~\ref{fig:cen_per}
shows that the shape of the $\pi^0$ spectrum changes significantly when
going from n+n to peripheral Pb+Pb collisions with $N_{coll} \approx 30$.

\section{Comparison to pQCD calculations}
\label{se:pqcd}
\begin{figure}
   \centerline{\includegraphics[height=7cm]{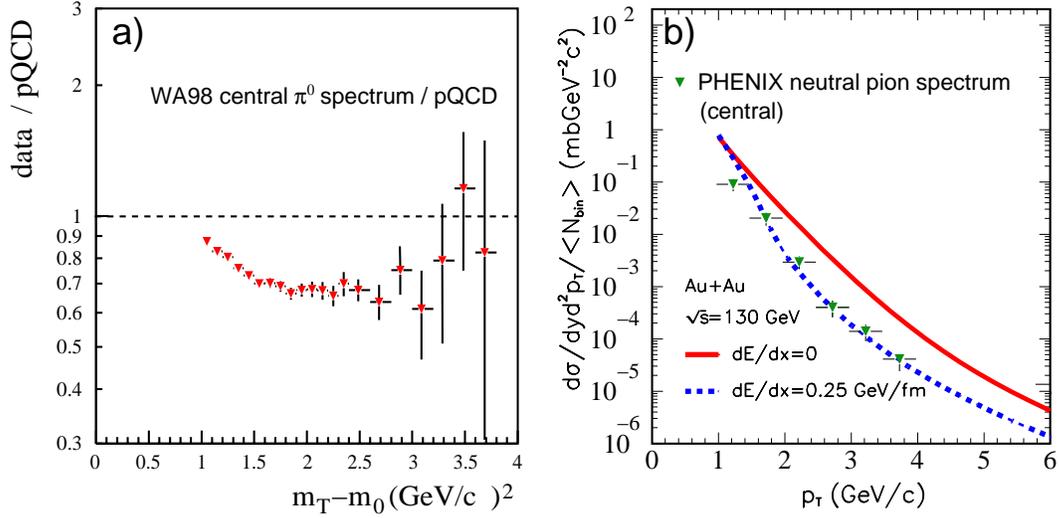}}
   \caption{\footnotesize
            Comparisons of neutral pion spectra measured in central 
            Pb+Pb collisions at the CERN SPS and in central Au+Au collisions
            at RHIC with pQCD calculations from X.N.~Wang 
            \cite{wang_sps, wang_rhic}. The left panel (a)
            shows the ratio of the WA98 central neutral pion spectrum to the 
            pQCD calculation. At RHIC energies (b) the differences between the
            standard pQCD calculation without parton energy loss and data
            are so significant that the spectra themselves can directly be 
            compared on a logarithmic scale.}
   \label{fig:pqcd}
\end{figure}

A pQCD calculation from Wang described in \cite{wang_sps} is compared to 
the central $\pi^0$ spectrum for Pb+Pb collisions at 
$\sqrt{s_{nn}}=17.3$~GeV/$c$ in Figure~\ref{fig:pqcd}a. In the range around 
$m_T-m_0 \approx 2.5$~GeV/$c^2$ the pQCD calculation overpredicts 
the experimental data by around 30\%. In a recent publication 
also L\'{e}vai et al. compare their pQCD calculations to 
WA98 data \cite{levai}. In agreement with Wang's calculation they 
find that their pQCD calculation predicts more neutral pions at high $p_T$
than actually measured. Thus, one can conclude that a possible parton energy
loss effect ("jet quenching") is not ruled out at SPS energies.

Figure~\ref{fig:pqcd}b shows a comparison of the PHENIX neutral pion
spectrum in central Au+Au collisions with a pQCD calculation from Wang
\cite{wang_rhic}. The standard pQCD calculation without parton energy 
loss effects clearly fails to describe the data points. By introducing 
a parton energy loss parameter a reasonably description of the data
can be reached. In \cite{wang_rhic, wang_wang} it is argued that 
when the expansion of the fireball is appropriately taken into account 
the energy loss of $dE/dX=0.25$~GeV/fm in an expanding system effectively 
corresponds to a much higher energy loss in a static system. 
This implies that the energy loss in the initial fireball produced in 
a A+A collision at RHIC is significantly higher than in cold nuclear matter.

\end{document}